\documentclass[3p]{elsarticle}

\usepackage[hyphens]{url}
\usepackage{xspace}
\usepackage{hyperref}

\newcommand{\oscp}{OSCP\xspace}
\newcommand{\ocpp}{OCPP\xspace}
\newcommand{\cpo}{CPO\xspace}
\newcommand{\emsp}{eMSP\xspace}
\newcommand{\ocpi}{OCPI\xspace}
\newcommand{\oicp}{OICP\xspace}
\newcommand{\ochp}{OCHP\xspace}
\newcommand{\ochpd}{OCHPdirect\xspace}
\newcommand{\openadr}{OpenADR\xspace}

\newif\ifignore

\begin{document}

\title{Security of EV-Charging Protocols}


%

\author{Pol~Van~Aubel\corref{cor1}\fnref{intdecl}}
\ead{pol.vanaubel@cs.ru.nl}

\author{Erik~Poll\corref{}\fnref{intdecl}}
\ead{erikpoll@cs.ru.nl}

\cortext[cor1]{Corresponding author}

\fntext[intdecl]{Declarations of interest: none}

\address{Digital Security group, Institute for Computing and Information Sciences, Radboud University%
\\
Toernooiveld 212, 6525 EC, Nijmegen, the Netherlands%
}

\begin{abstract}
	The field of electric vehicle charging involves a complex
	combination of actors, devices, networks, and protocols. These
	protocols are being developed without a clear focus on security.
	In this paper, we give an overview of the main roles and protocols
	in use in the Netherlands. We describe a clear attacker model
	and security requirements,
	show that in light of this many of the protocols have
	security issues, and provide suggestions on how to address these
	issues.
	The most important conclusion is the need for end-to-end security
	for data in transit and long-term authenticity for data at rest.
	In addition, we highlight the need for improved authentication of
	the EV driver, e.g. by using banking cards.
	For the communication links we advise mandatory use of TLS,
	standardization of TLS options and configurations, and
	improved authentication using TLS client certificates.
%
\end{abstract}


\maketitle

\section{Introduction}
\label{sec:introduction}
Similar
to how petrol-powered cars require an omni\-presence of gas
stations, electric vehicles (EVs) require an infrastructure of charging stations. However,
where a transaction at most gas stations is a matter of paying on-premises without
prior existence of a contract between the supplier of petrol and the
driver of the vehicle, the charging infrastructure of electric
vehicles is very contract-oriented. Billing is usually performed
monthly, on a post-paid basis. For this to work, there needs to be a
system to track the charge sessions of the EVs.

Although to the outsider the charging infrastructure may simply seem
like a series of electrical outlets to hook cars up to the electric
grid, behind the scenes we find a more complex picture. Charge points are
connected to back-end systems of \emph{charge point operators
(\cpo{}s)}, which in turn
communicate with \emph{e-mobility service providers (\emsp{}s)}.
Cars and charge points
communicate to inform each other about their capabilities and
restrictions. For all these interactions, protocols have been designed
to exchange the required data.
Broadly, there are two categories of data that we can distinguish:
\begin{itemize}
	\item \emph{billing-related data}, such as
		reports of meter values before and after a charge session, and
	\item \emph{control-related data}, such as instructions to a
		charge point of how much current it is allowed to draw.
\end{itemize}
The control category is important for availability.
Disrupting or corrupting that data attacks the stability of the power grid,
can trigger physical protections to prevent overcurrent,
etc. \cite{acharya2019}

Another, related, distinction is whether the data is
considered personal data under the GDPR \cite{GDPR}.
Although it might seem evident that billing-related data is 
personal data and control-related data is not, the distinction is not
necessarily this straightforward. E.g., control-related data may carry
information about the behaviour of the battery being charged. Even
though this data has no direct identifiers, it could have sufficient
information to accurately identify the specific battery, i.e. the
car, being charged. We therefore advise a conservative mindset with
regards to data sharing and data use: only share that information that
is actually necessary to perform the task at hand; and encrypt all
data, not just the data that has been determined beforehand to be
personal data. This helps to ensure privacy by default and by design.

With so many data flows and so many actors, the security of the
ecosystem may suffer from a weak link anywhere in the chain.
In this paper we analyse the security aspects of this
ecosystem.
In Section~\ref{sec:overview} we provide an overview of
actors and their roles in the EV-charging infrastructure in the
Netherlands, and introduce the protocols that are currently in use to
facilitate communication between them.
Section~\ref{sec:security} classifies the attackers and describes
security requirements for the EV-charging infrastructure.
Section~\ref{sec:securityaccesscontrol} provides an analysis of the security
issues we see with access control, and Section~\ref{sec:securitydata}
does the same for security of the communicated data. Both sections also
suggest improvements to the current situation.
Although we will not focus on the privacy aspects of EV charging in
our security analysis, they do influence some of our recommendations,
so we will briefly discuss them in Section~\ref{sec:privacy}.
In Section~\ref{sec:futurework} we discuss some ideas for future work.
Finally, in Section~\ref{sec:conclusion} we summarize our findings.

This paper builds on earlier work \cite{CAGS2015} by considering
additional protocols, presenting a more detailed security model, and
exploring security issues more in-depth.



\section{The Dutch EV-charging landscape}
\label{sec:overview}

This section fixes our terminology and
describes the various roles and protocols
in the EV-charging ecosystem that we need to distinguish.
We group several components that are in reality considered separate;
e.g. we will not distinguish between an EV and its embedded
communication controller. We note that this is but one way
to view a complex market, but we believe it to be sufficient to
understand the security implications.

\subsection{Roles}
The most important roles in the EV-charging ecosystem are:
\begin{enumerate}
	\item The \emph{\cpo (Charge Point Operator)} operates and
	maintains charge points. \cpo{}s play an important role in the EV
		market, as they interact with the DSO (see below) and the
		\emsp{}s\footnote{Different documents use different terms for
		similar roles. E.g., ISO~15118 calls the role of \cpo
		\emph{Electric Vehicle Supply Equipment Operator}
		and the role of \emsp \emph{Electric Vehicle Service Provider}.
		An additional complication is the
		custom to indicate car manufacturers as \emph{Original Equipment
		Manufacturer (OEM)}. But what an OEM is differs depending
		on context -- to a \cpo an OEM might as well be the
		manufacturer of the charge points, rather than the cars. We
		therefore refrain from using this term.}.

	\item The \emph{\emsp (E-Mobility Service Provider)} (re)sells
		the electricity to EV drivers. The \emsp has contracts with
		EV drivers and takes care of billing them.

		The role of \emsp can be fulfilled by specialized parties, but
		can also be fulfilled as a secondary activity by an existing
		actor. For example, if
		the EV driver pays directly for a charge session
		with his credit card, then the credit card provider
		takes on the role of \emsp.

	\item The \emph{DSO (Distribution System Operator)} manages the
		regional electricity grid and is responsible for its
		stability and reliability. They also usually operate the
		metering equipment for the grid connection of the charge points.


	\item The \emph{Clearing House} offers a platform to
		exchange data between \cpo{}s and \emsp{}s in a standardized way,
		possibly across national borders.
		There will be many \cpo{}s and \emsp{}s, and a single \emsp can
		have contracts with many \cpo{}s to allow its clients to use the
		charge points of these \cpo{}s. Rather than making point-to-point
		connections everywhere, parties can use a clearing house to facilitate
		the necessary interactions.

	\item The \emph{Electricity Supplier} provides the
		electricity consumed at a charge point.
		There
		are a few options for contracting the electricity supplier.
		The two most obvious are:
		\begin{itemize}
			\item the electricity supplier has a contract with the
				\cpo, who in turn bills the \emsp{}s for the incurred use;
				and
			\item the \emsp has a contract with an electricity
				supplier, and is billed directly by them.
		\end{itemize}
		An issue in the latter case is how the \cpo makes money on its
		services -- one solution is for the \cpo to bill the \emsp for
		use of the charge point.


	\item The \emph{CPIO (Charge Point Infrastructure Operator)} is
		typically a vendor or manufacturer of charge points and
		performs some maintenance, such as updating firmware, on
		behalf of the \cpo.  In some situations the actual maintenance
		is performed by the \cpo itself, i.e., updates are sent by the
		CPIO to the \cpo and the \cpo takes care of them, but in other
		cases it is done directly by the CPIO.

	\item The \emph{Car Manufacturer} that manufactures cars
		compatible with the EV-charging infrastructure.
\end{enumerate}
These roles need not be performed by different actors: a
DSO may operate charge points, i.e. act as \cpo,
and one actor could be both \cpo and \emsp;
Tesla is a car manufacturer that also acts as
\cpo (Tesla fast charge points) and \emsp (Tesla fast charge credits).
However, there may be legal constraints on which roles a given actor
may play. In particular, competition laws may restrict which roles
a DSO, as a monopolist, is allowed to play.
In the Netherlands there have
been court cases about whether DSO-owned \cpo{}s can also sell
electricity and thus also act as
\emsp \cite{NuonVsAllego}.
Similarly, a car manufacturer that is the \emph{only} possible \emsp for its
customers might be accused of anti-competitive behaviour.

\medskip

In addition to the roles listed above, we highlight two more:
\begin{itemize}
	\item \emph{Value-Added Services} are providers of additional
		services not previously mentioned. E.g., a \emph{Navigation
		Provider} such as Google, TomTom, or Garmin may offer services
		for EV drivers to find available charge points.  A \emph{Parking
		Spot Operator}, e.g. a parking garage, might collaborate with
		a \cpo to offer charge points.  These parties fall outside of
		the scope of this paper, but it should be noted that our
		security concerns may extend to the data exchanged with them
		and the protocols used for that.

	\item Finally, there are \emph{Industry Consortia}
		that cut across roles to bring parties together in an effort
		to improve collaboration. Examples are the NKL
		organization\footnote{\url{https://nklnederland.nl/}},
		ElaadNL\footnote{\url{https://www.elaad.nl/}}, and the Open
		Charge
		Alliance\footnote{\url{https://www.openchargealliance.org/}}.
		One major activity these consortia undertake is the
		standardization and promotion of protocols.
\end{itemize}

\subsection{Protocols}

\begin{figure}[h]
	\centering
	\includegraphics[width=.55\textwidth]{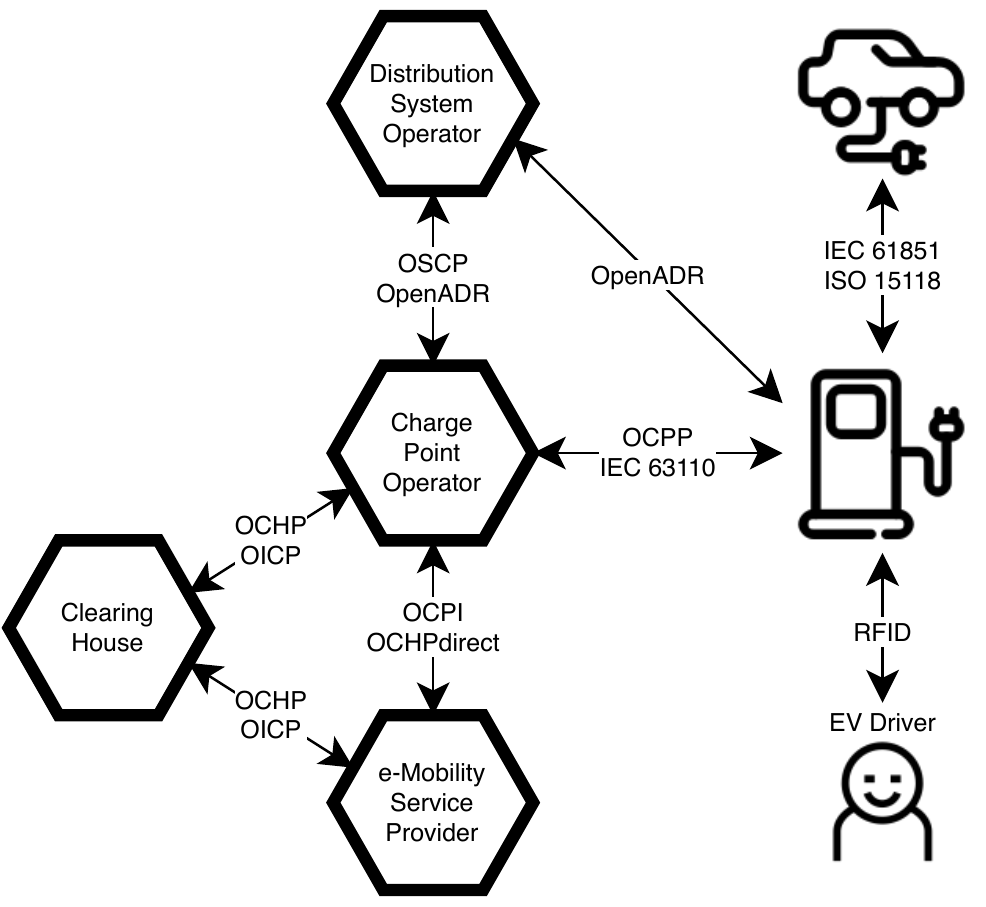}
	\caption{Protocol landscape of the Dutch EV-charging infrastructure}
	\label{fig:protocoldiagram}
\end{figure}

The communication infrastructure between the various parties needs to
facilitate the following processes:
\begin{enumerate}
	\item Authorizing an EV to charge. This involves
		identification and authentication of the EV and/or the
		EV driver.

	\item Billing of EV drivers and billing between market parties.

	\item Management of the charge point infrastructure. This
		includes detecting, registering, and reporting EVs that
		negatively impact charging service.
	\item Influencing EV charging behaviour to integrate better in
		the power grid. There are two main aspects to this:
		\begin{enumerate}
			\item \emph{Congestion management} is mainly concerned with not
				overloading the grid. E.g. if several charge points
				share a grid connection, their combined load should
				not overload the connection. This may require actively
				influencing the charging behaviour of the attached EVs,
				charging them all at a lower rate or charging them
				sequentially.

			\item \emph{Demand-supply balancing} involves
				influencing the demand to counterbalance fluctuating
				supply (esp. from wind and solar), by e.g. charging
				more or fewer cars, influencing their charge speed,
				or even by discharging cars, effectively using car
				batteries as energy storage for the grid.
		\end{enumerate}
		Although the industry does not appear to have settled on a single
		agreed definition of the term ``smart charging'', the
		definitions we have encountered are all variations on one or
		both of these aspects.
\end{enumerate}
The protocol landscape for this is still in flux. For each of
the connections between actors, different protocols exist, in various stages of
standardization. Because EV charging is a relatively young field,
extensions and new protocols are constantly being developed.
Figure \ref{fig:protocoldiagram} gives an overview of the protocols
currently available for communication between actors, and these
protocols are discussed below. We will not explain their
functionality in depth, since we are mostly interested in the security
implications and guarantees. For a more extensive and
in-depth review of the functionality of these protocols, we refer to
\cite{elaadprotocols}.

\subsubsection{Communication between EV and Charge Point}
Charge Points provide one or more sockets where EVs can be charged.
The EV and charge point communicate over the cable that is used for
charging.

\begin{itemize}
	\item \emph{IEC~61851} \cite{IEC61851}. This protocol is also
		known as the Mode~3 protocol.
		It is supported by practically all
		currently available EVs. Communication between the EV and the
		charge point is minimal, using a basic pulse-width
		modulation protocol that ensures that charging happens
		without technical problems.

	\item \emph{ISO~15118} \cite{ISO15118-2}.
		This
		is the intended successor of Mode~3.  Unlike Mode~3, it is an
		extensive protocol for communicating information between
		charge point and EV. It introduces an authentication
		mechanism called Plug-and-Charge to identify and authenticate
		the EV. It also adds the possibility for the EV
		to sign records of meter readings, so ISO~15118 also
		involves data and functionality that is of interest for the
		\cpo and the \emsp.
\end{itemize}
Since Mode~3 is a very basic protocol that only communicates to
establish the technical parameters of a charge session,
it is not considered in the remainder of this paper -- we only mention
it for completeness' sake.

\subsubsection{Communication between Charge Point and \cpo}
A charge point has a communication link, for
instance a GPRS connection, to the back-office of the \emph{Charge
Point Operator (\cpo)}.

\begin{itemize}
	\item \emph{\ocpp}. The Open Charge Point Protocol
		\cite{OCPP} is the dominant protocol in use.
		It standardizes the communication between the charge
		point and the \cpo. It allows
		back-ends and charge points of different vendors to
		communicate, simplifying operations and preventing vendor
		lock-in. As part of that, \ocpp also allows for remote
		maintenance of charge points by the \cpo or CPIO through
		monitoring and firm\-ware updates. It also offers features
		needed for influencing charging behaviour, notably limiting the maximum
		capacity that a charge point can deliver to an EV in a certain
		time slot.
		\ocpp has seen several revisions, and the security aspect of
		\ocpp has significantly changed from version 1.6 to version
		2.0. Since OCPP 1.5 and 1.6 are still widely used, our
		analysis distinguishes between the versions where
		applicable.

	\item \emph{IEC~63110}. This is an effort by the IEC to arrive
		at a standardized protocol that fulfils the same role as
		\ocpp.  \ocpp version 2.0 was one of its foundational inputs,
		but we have not had a chance to see drafted documents, so
		we cannot analyse whether our security requirements from
		Section~\ref{sec:attackermodel} are satisfied.
\end{itemize}

\subsubsection{Communication between \cpo, \emsp, and Clearing
House}

\begin{itemize}
	\item
		\emph{\ocpi}.
		The Open Charge Point Interface \cite{OCPI} is a JSON-based
		protocol intended to enable EV drivers to use the charge
		points of many different \cpo{}s
		without requiring a third party such as a clearing
		house.

	\item
		\emph{\ochp}.  The Open Clearing House
		Protocol \cite{OCHP}
		and its extension OCHPdirect are a set of SOAP-based protocols
		to facilitate
		connections between a central clearing house, \emsp{}s, and
		\cpo{}s. \ochpd enables peer-to-peer connections, similar to
		\ocpi, but does require a clearing house to negotiate the
		connections.

	\item
		\emph{\oicp}. The Open InterCharge
		Protocol \cite{OICP}
		is another JSON- and SOAP-based
		protocol facilitating clearing house communication, at the same
		level as \ochp.
\end{itemize}

\subsubsection{Communication between \cpo and DSO, or Charge Point and
DSO}

To ensure stable operation of the grid when faced with high-capacity
charge points, the DSO needs to be able to inform the \cpo
about the capacity and supply \& demand state in this moment. There are
two protocols in use for this:
\begin{itemize}
	\item \emph{\oscp}. The Open Smart Charging Protocol \cite{OSCP}
		enables negotiation between a DSO and \cpo{}s. The DSO creates a
		supply \& demand forecast on 15-minute intervals.
		The \cpo is then informed of its allotted capacity and
		the remaining spare capacity, but it can negotiate for more or
		less capacity.
		The \cpo then creates a charge plan for the charge points,
		specifying the limit of the power they can supply per time
		slot, and transmits this to the charge points using e.g.
		\ocpp.

	\item \emph{\openadr}. Open Automated Demand Response
		\cite{OpenADR} is a protocol developed by the primarily US-based
		\openadr Alliance, for automated demand response and dynamic
		price communication.
		It provides more direct options for a DSO to manage equipment,
		e.g. giving the DSO the ability to turn equipment off directly if
		demand exceeds supply.
\end{itemize}

\subsection{Protocol layering \& data types}
For our security analysis in the following Sections, it is important
to understand protocol layering and how the underlying protocol layers
relate to security, and to understand the distinction between data in
transit and data at rest.

All the aforementioned protocols are application protocols, i.e.
they are the uppermost layer of a layered protocol model (e.g. both
the OSI model and the IP model
have an application top layer \cite{OSIModel,RFC1122}).
They define a particular set of allowed
messages, with semantic meanings in the application domain.
For formatting these messages, application protocols are often based
on other standards.
The two most common choices for message formatting in this ecosystem
are SOAP/XML and JSON. Notably, both \ocpp and \oicp have taken the
decision to move from XML to JSON; in contract, ISO~15118 is a
relatively new protocol specified
for XML. Ideally, the information transported by one protocol should
easily be transferable in another protocol, so conversion between
message formats is required.

Application protocols can specify the use of other, underlying,
protocols (e.g.  TLS, TCP, and IPv4) to transport the messages:
transport-layer protocols.
Transport-layer protocols run between two directly communicating
hosts, and are usually unaware of the semantic meaning of messages. An
application-layer protocol may be specified with message
or data forwarding in mind, either via other application-layer
protocols or by using multiple transport-layer hops. This means that
there may be intermediary parties between the communicating parties.

It is not required for an application-layer protocol to specify every
detail of its underlying protocol stack. However, as we will argue in
Section~\ref{sec:tlsspec}, if
the \emph{security} of the underlying protocols is relevant for the
security of the application-layer protocol, then the application-layer
protocol should specify the requirements as detailed as possible,
preferably by mandating and limiting the allowed protocols and
configurations for these protocols.

Finally, we need to distinguish
between \emph{data in transit} and \emph{data at rest}. Data in
transit is the data being communicated by protocols between
endpoints. Data at rest refers to data stored for (eventual) processing.
One example of data at rest are stored Charge Detail Records (CDRs), which are
descriptions of concluded charging sessions and are used to bill
actors. Data at rest has usually, at some point, been data in transit.

\section{Attacker model \& security requirements}
\label{sec:security}
In this section, we first classify attackers that might
attack the EV-charging ecosystem based on their capabilities.
Then, in Section~\ref{sec:securityrequirements}, we lay out security
requirements for the EV-charging landscape, based on the processes and
roles from Section~\ref{sec:overview}, and clarify how these requirements
protect against the attacker classes introduced. In
Section~\ref{sec:srlimitations} we discuss some limitations of our
approach.

\subsection{Attacker model}
\label{sec:attackermodel}
A clear definition of the types and capabilities of attackers is
critical for any security analysis.
We can broadly categorize attackers in three distinct categories:
\begin{enumerate}
	\item Physical system attackers: these use physical
		access to compromise a single system. This attacker, if
		successful, becomes an attacker of the second type.

		The systems most susceptible to physical attacks are the
		charge points and the EVs themselves, because these are
		located in the field. They can be attacked by
		their owners and any interested passer-by.
		Practical attacks on charge points used in the field have
		been demonstrated in the past \cite{schwarzladen}.

	\item Malicious systems, a.k.a. end-point attackers:
		legitimate systems that, through compromise by an attacker or
		other means, have now become adversaries in the EV-charging network.
		This is not limited to just the charge points or EVs, but also includes
		the IT systems run by e.g. \cpo{}s and \emsp{}s to fulfil their
		roles.

		It should be noted that malicious or incompetent insiders also
		give rise to malicious systems.

	\item Network attackers: attackers that attack the network
		traffic.
		Network attackers can usually be stopped by proper
		authenticity and confidentiality mechanisms.
\end{enumerate}

\subsection{Security requirements}
\label{sec:securityrequirements}
Data exchanged between roles is intended to facilitate their business
processes. There need to be assurances on this data.
Consider, for example, the following scenarios, which are not all
between parties that communicate directly:
\begin{itemize}
	\item An \emsp wants to ensure that only \cpo{}s it has contracts
		with can push data to its systems.
	\item A charge point wants to ensure that a connecting EV is
		allowed to charge. A special case of this is when the charge
		point is currently not connected to the internet.
	\item A \cpo wants to ensure that only charge points it owns can
		connect to the communication interface for its charge point
		protocol.
	\item An \emsp wants to ensure that a \cpo cannot deny having sent
		them a particular Charge Detail Record (CDR).
	\item A \cpo wants to ensure an \emsp cannot falsify CDRs.
	\item An EV wants to ensure that the tariff table it receives
		comes from the \emsp its driver has a contract with.
	\item An \emsp does not want to show the actual tariff it
		negotiates with the EV to the \cpo.
\end{itemize}
The final two points bear some clarification.
As part of ISO~15118, the EV can negotiate the charge speed based on a
tariff table that is receives from the charge point. However, these
rates are ultimately provided by the \emsp, forwarded by the \cpo and the
charge point to the EV. The accuracy of this table and the
rate the EV decides to use directly influences the billing process.
In the current system, the EV and \emsp have to trust the \cpo to
pass the traffic going in either direction without changes, and not to
use the information contained within in an anti-competitive manner.
E.g. the \cpo could only send the rates that result in the
highest profit for the \cpo to the EV, ensuring that the car selects
one of those rates. The \cpo could also simply pretend to the \emsp
that a high rate was selected, effectively
making the \emsp pay for services not provided.
Another risk is that the \cpo simply records all the tariff
negotiation, and then e.g. sells that information to another \emsp.

These scenarios are of course not exhaustive, but they serve to
illustrate the need for the security requirements below.
We propose nine security requirements (SRs), grouped in five
categories:

\medskip
\subsubsection{Access control for the charging infrastructure}

This category ensures legitimacy of charging EV drivers.

\smallskip
\paragraph*{SR 1a) Authentication of the EV driver or EV}
E.g. accomplished by using a credential such as a smart card, or a
contract certificate embedded in the EV.
	
\smallskip
\paragraph*{SR 1b) Authorization to charge}
An authenticated EV or EV driver needs to be authorized to charge at a
charge station.

\smallskip
\paragraph*{SR 1c) Availability of charging}
An EV driver should not wrongfully be denied charging. It is important
to keep in mind that charge points are not necessarily connected to
back-end systems with a reliable connection, so a charge point may not
always be online.
Another concern could be that even though an EV driver should not be
allowed to charge at a particular charge point, they should be
provided with a minimum charge to ensure they do not get stranded
somewhere. This is a business decision, not necessarily something that
should be codified in protocols, and will not be examined further in
this paper.

\medskip
\subsubsection{Strong authentication of systems}
This category ensures legitimacy of the communicating systems.

\smallskip
\paragraph*{SR 2a) Strong authentication of servers to clients}
A client connecting to a server needs to be able to verify it is
talking to a legitimate server.

\smallskip
\paragraph*{SR 2b) Strong authentication of clients to servers}
A server being connected to by a client needs to be able to verify the
client is legitimate.

\medskip
\subsubsection{Secure transport links}
A conservative choice is to always at least ensure security
(authenticity\footnote{
Note that we distinguish between authentication of the EV driver, as
part of access control, and authenticity of data as part of security
for the communicated data. Although these concepts are related, we
treat them separately because the authentication of the EV driver is
required for authorization, whereas the authenticity of data is
required throughout the ecosystem.} and confidentiality) for
point-to-point transport links.  Secure transport links ensure that no
network attackers, i.e.  attackers of type 3, can read or modify data
being exchanged between two directly communicating parties.

\smallskip
\paragraph*{SR 3) Use of TLS on every communication link}
Although there exist other options of securing transport links, e.g.
Virtual Private Networks, we believe it is desirable to standardize on
a single, universally applicable, technology, and TLS is that
technology. Therefore, we make this choice explicit in these
requirements.

\medskip
\subsubsection{End-to-end security for the data in transit}
Whereas SR~3 only protects against attackers of type 3, end-to-end
security requirements SR~4a and 4b
also protect against attackers of type 1 and 2 that
are on a point between two communicating parties.

To understand the difference, consider that different actors
in different roles may be forwarding data between
communicating parties.  E.g., if the EV
is charging at a charge point, communication with the \emsp is
proxied by the \cpo. Even if secure transport links
between EV, charge point, \cpo, and \emsp exist, then the EV
and \emsp must trust the \cpo and charge point not to modify
the data being forwarded. If an attacker of type 2 has
managed to compromise the \cpo or the charge point,
secure transport links do nothing to protect the data.
Therefore, TLS cannot satisfy these security requirements.

\smallskip
\paragraph*{SR 4a) End-to-end authenticity of application-layer data}
This data includes e.g. firmware upgrades and CDRs.
It needs to be verifiable that data is indeed produced by
the party that is expected to produce it.

\smallskip
\paragraph*{SR 4b) End-to-end confidentiality of application-layer data}
This ensures that data is only readable for intended recipients. This
requirement stems from business requirements as well as privacy
requirements, because it provides:
\begin{itemize}
	\item Confidentiality of sensitive business data,
		such as charge tariff lists.
	\item Privacy of the EV driver. Information transmitted
		may include e.g. the location where an EV driver
		was at a certain time. This is personal data as
		defined under the GDPR \cite{GDPR}. Legal requirements
		on the handling of personal information therefore
		apply. Though not our primary focus, we will briefly
		discuss privacy of the EV-charging infrastructure in
		Section~\ref{sec:privacy}.
\end{itemize}
	
\medskip
\subsubsection{Non-repudiation for data at rest}
In our attacker model we assume parties may act maliciously,
and therefore we cannot assume the long-term authenticity of the data
used in e.g. the billing process, i.e. the data at rest.
SR~4a and 4b do not prescribe authenticity guarantees for data at
rest, so we need an additional requirement.

\smallskip
\paragraph*{SR 5) Non-repudiation of (billing-related) application-layer data}
This prevents a party from denying having produced a (billing-related)
message or commitment. This auditable trail of messages can then be
used to resolve disputes.

Note that this is a stronger requirement than SR~4a, because
non-repudiation requires authenticity guarantees, but solutions that
provide authenticity do not necessarily provide non-repudiation.
If there are authenticity or confidentiality guarantees for data at rest,
provided by the original producer of the data, then these typically
also hold \emph{end-to-end} for that data in transit. E.g. if an application
defines a digital signature mechanism on messages for long-term authenticity
guarantees, these signatures can be checked upon initial
receipt of these messages, and appropriate measures
can then be taken if the signatures fail to verify.
Therefore, a mechanism to satisfy SR~5 may also be used to satisfy
SR~4a.

One particular instance where failure to satisfy SR~5 is
worrying is the generation and storage of Charge Detail
Records (CDRs).  The \ocpi standard requires CDRs to be
immutable objects. \cpo{}s generate CDRs and send them to
\emsp{}s. After the \cpo sends it, neither \cpo nor \emsp is
supposed to change the CDR, but without authenticity and
non-repudiation, neither party can verify or prove that
immutability.

We note that SR~4a, SR~4b, and SR~5 must be implemented in such a way
that privacy requirements from the GDPR can be satisfied, which we will
explain in Section~\ref{sec:secreqsgdpr}.

\subsection{Impact \& limitations}
\label{sec:srlimitations}
The charging infrastructure represents a potentially very large
dynamic load on the grid.
The European power grid is designed to be able to cope with imbalances
of 3 gigawatt \cite{entsoe-dispersed}.
We do not have exact figures, but from private communication we
understand that the potential load from the EV-charging infrastructure
is likely to exceed this threshold within the next decade.
Such a load may accidentally or intentionally be
manipulated to destabilize the grid \cite{acharya2019}.  The only
contribution w.r.t. this aspect we can make in this paper
is the observation that we
should minimize the possibility that the EV-charging systems are
manipulated by bad actors. To that end, our listed security
requirements are paramount.

Finally, the security requirements we listed offer no solution for the case
where an attacker of type 1
or 2 has subverted one of the sending or receiving parties in a
communication:
an attacker that can pose as a legitimate participant in the protocols
can use all the features provided by those
protocols. If e.g. a charge point has a remote off-switch that a CPO
can trigger, then an attacker that can pose as that CPO
could try to use it. Or, if a large amount of energy can be
reported as having been transferred from the car to the grid, an
attacker might get reimbursed for the energy. Preventing or detecting
abuse of features by attackers that can pose as legitimate actors may
be assisted by the authenticity guarantees from SR~4a and SR~5, in the
form of audit logs. However, the implementation and use of monitoring
\& logging is external to the protocol definitions, and is therefore
out of scope for this paper.


\section{Security issues in access control}
\label{sec:securityaccesscontrol}
As mentioned in Section~\ref{sec:securityrequirements}, there should
be access control for the infrastructure, consisting of SR~1a,
authentication of the EV driver, and SR~1b, authorization to charge.
The major issue with this is the specific way in
which RFID cards are currently used to identify the EV driver,
which we will explore in the first part of this Section, and then
we suggest some improvements.


\subsection{Using UIDs for authentication of the EV driver or EV}
\label{sec:evdriverauth}
At public charge points drivers are authenticated through the use of an
RFID card. As already mentioned in~\cite{CAGS2015}, every customer is
identified using only the card's UID that is transmitted plaintext
through the air.
We will refer to this mechanism as the \emph{weak UID
method}.
This can hardly be called authentication, because transmitting the
UID is sufficient to be authenticated as that UID.

The UID is always broadcast as part of communication with the card.
This means that learning the UID is trivial if an attacker has access
to the card: they can simply read the information using a standard
NFC-enabled phone.  With specialized equipment it
is also possible to eavesdrop on the communication between the card
and a charge point. This may be possible at a distance of several metres
\cite{Engelhardt2013,Habraken2015}. However, similar to ATM skimming
devices, an attacker could simply attach their eavesdropping equipment
to the charge point.
Then, when the attacker has a valid UID, they can simply configure it on
a card with a configurable UID, or spoof it with e.g. an NFC-enabled
mobile phone \cite{francis2009}.

The RFID cards currently used are mainly MIFARE Classic cards
\footnote{\url{https://www.mifare.net/en/products/chip-card-ics/mifare-classic/}}.
These cards are capable of a stronger authentication method, using a
challenge-response protocol, but even then this authentication method
is very weak, as the proprietary cryptography used here has been
broken \cite{garcia2008-1,meijer2015}.


However, we note that even though cloning cards is so easy,
this does not necessarily mean
there will be a problem in practice.
The MIFARE Classic has been used in
public transport in London (Oyster) and the Netherlands
(OV-chipkaart), and in both cases this has not caused significant
amounts of fraud in the past ten years.
In the case of EV charging, the risk to the fraudster is similar:
being caught red-handed using a cloned card while still hooked up to a
charge point, so it may turn out that we will not see a significant
amount of fraud here either.
Therefore, any move to better mechanisms as suggested
below
may be driven more by
technological advancements, or advantages in aspects other than security
such as the ease of Plug-and-Charge, rather than any immediate need
due to fraud.


\medskip
\subsubsection*{Security improvement: challenge-response authentication}
\label{sec:disevdriverauth}
Any improved authentication mechanism would need to use a
challenge-response mechanism, instead of just reading the UID of an
RFID card.
Such a challenge-response mechanism can be implemented in various
ways:
\begin{enumerate}
	\item Charge points need a shared symmetric master key with
		the cards, or
	\item Charge points need to know the asymmetric public key of an
		authoritative certificate to be able to authenticate the
		cards, or
	\item Charge points always need to be online with a direct
		connection to off-load the verification to the issuing party.
\end{enumerate}
We currently have no clear indication that
challenge-response authentication, in any of these forms, is
implemented anywhere in the EV-charging ecosystem.
Option 3, the always-online option, would potentially conflict with
SR~1c, which means keys need to be distributed to the charge points.
Option 1 would require distributing symmetric
shared keys to all the charge points in the field, which, as mentioned
in Section~\ref{sec:attackermodel}, is vulnerable to physical
attackers. If a symmetric key were to leak, the entire system would
break down. Therefore, we believe the best choice to be option 2, the
asymmetric option.

ISO~15118 introduces precisely such an asymmetric cryptographic option:
Plug-and-Charge.
Instead of the driver using an RFID card, Plug-and-Charge enables the
EV itself to identify and authenticate to the charge point, via the
charge cable.
This effectively replaces authentication of the EV driver with
authentication of the EV.
Plug-and-Charge uses X.509 contract certificates with standardized
certificate profiles, which are used to sign certain messages on the
application layer. The public key
infrastructure required for this is discussed in
Section~\ref{sec:distlsspec}.
However,
ISO~15118 also provides for External Identification Means (EIM).
This means that if an EV does not support Plug-and-Charge, other
mechanisms like RFID cards can still be used. Therefore, these
mechanisms will exist side-by-side, and we should also use an improved
card mechanism.

When deciding on that mechanism, we should bear in mind that
the EV-charging ecosystem is not the first to have to solve
this problem of authentication of moving actors using cards.
For example, the banking sector has a long history of providing working
authentication across multiple parties, in multiple locations (ATMs
and payment terminals). Current contactless banking cards are based on
the EMV standard. The EMV standard does not only facilitate secure
payments; it is also possible to only authenticate the card to the
reader using asymmetric cryptography \cite{EMVbook2}. This is the basis for
contactless bank card authentication systems such as a trialled
replacement for the Dutch public transport card. The EV-charging
ecosystem could also use this EMV card authentication method, provided
the card readers on the charge points are upgraded to use EMV.


Although initially it may seem that the use of EMV card authentication
would require Payment Card Industry certification
\footnote{\url{https://www.pcisecuritystandards.org/pci_security/maintaining_payment_security}},
we understand from private communication with the payment sector that
that is not necessarily the case.
EMV is an open standard, the public keys required to authenticate
the cards are publicly available\footnote{E.g.
\url{https://www.eftlab.com.au/knowledge-base/243-ca-public-keys/}},
and no communication with the international payment system is required
to perform this authentication.
Therefore, as long as EMV card authentication is only used for driver
identification and authentication, PCI certification of
implementations is not required.
Of course, if the charge points also have the possibility of actually
paying by card
directly on-premises a certified terminal is already present. This
terminal could then also be used for EMV card authentication.

Another option that uses asymmetric cryptography is the use of
NFC-capable smartphones, performing the same authentication steps as an
EMV banking card, as Apple Pay and Google Pay do.

We think that these options should be sufficient to provide
strong EV-driver authentication.
Alignment with EMV also means that many standard off-the-shelf
solutions already exist, and no custom solution has to be built.
However, if the industry still wants a custom-built RFID system, there
is another obvious option: alignment with ISO~15118 by running the
Plug-and-Charge authentication methods of ISO~15118 on the RFID card
itself. As we will see in Section~\ref{sec:dise2e}, this has the added
benefit of providing stronger guarantees for SR~4a and SR~5 in the
case where cars do not support ISO~15118.
Again, a custom smartphone app using NFC communication could also be
used for this.

A challenge-response protocol based on public key cryptography would
be required regardless of the precise implementation, and would
probably end up looking a lot like the card authentication of EMV or
Plug-and-Charge authentication of ISO~15118. In any case, it would
likely still involve an upgrade of many existing card readers.

\section{Security issues for the communicated data}
\label{sec:securitydata}

There are multiple categories of security requirements for the data.
Recall from Section~\ref{sec:securityrequirements}:
\begin{itemize}
	\item Secure transport links (SR~3)
	\item End-to-end security (SR~4a \& SR~4b)
	\item Non-repudiation for data at rest (SR~5)
\end{itemize}
For all of these, authentication of the communicating systems is
required (SR~2a, strong authentication of servers to clients; and SR~2b,
strong authentication of clients to servers).
One issue we see here is the use of
static tokens to identify \& authenticate these systems,
which we will explain in Section~\ref{sec:systemauth}.

TLS is currently used to provide some of these authentication,
authenticity, and confidentiality requirements.
However, this is underspecified in many protocols, which we
explain in Section~\ref{sec:tlsspec}.

Finally, in Section~\ref{sec:e2e} we suggest improvements to the
current situation where, even if proper
authentication of systems is present and even if extensively
specified TLS is used, neither end-to-end authenticity,
nor end-to-end confidentiality, nor non-repudiation are provided by
the current versions of the protocols.

\subsection{Authentication of
systems using static credentials}
\label{sec:systemauth}
There are two main ways to use TLS:
\begin{itemize}
	\item with only server certificates, where servers do not
		authenticate the clients, as is usual for e.g. websites.

	\item with server and client certificates, where server and
		client use the same authentication mechanism to mutually
		authenticate each other.
\end{itemize}
All protocols we considered satisfy SR~2a when they use TLS with server
certificates. In that case, the server is
authenticated as web servers usually are, i.e. by being at a
certain URL and having a valid TLS server certificate for that URL.
However, not all protocols make TLS mandatory. In
particular \ochp, \ocpi, and \ocpp~1.5 and 1.6
leave TLS optional. In the absence of TLS the client cannot
authenticate the server at all.

For client authentication, the situation is more complex.
Several protocols -- in particular \ocpi and \ocpp in all but its
highest security profile --
use some form of static credential as a secret to identify and
authenticate the client to the server. \ocpi uses random bitstrings,
and most versions of \ocpp use username/password combinations.
These function fundamentally in the same way.
These credentials are \emph{shared} and \emph{static}:
all requests carry the same credential until it is updated to a new one.
For the initial setup of the protocols these credentials are generated
by the participants and sent to each other out-of-band, e.g. via
e-mail. After this setup, the credentials can be updated in-band using
the protocols themselves.
These credentials are included in each request.
Such a mechanism could be considered secure if the initial
distribution is done securely, and if TLS is used on the
transport layer. However, TLS is not yet mandatory for all protocols,
which exposes the secret to a higher risk of leaking, and therefore
this mechanism does not currently satisfy SR~2b.

The main issue is that possession of the secret is sufficient to pose
as a legitimate client. The risk of leaking the secret should be
minimized, e.g. by using it to
derive session secrets in a deterministic way or via some
challenge-response protocol, so that the secret itself only needs
to be transmitted when it is updated.
However, as we will see below, there are standardized mechanisms in
TLS that can replace these static credentials, so building an improved
version with challenge-response seems wasted effort.



We also see a certain asymmetry in these setups: server authentication
to the client is done on the transport layer using TLS and client
authentication to the server is done on the application layer using
static credentials. This asymmetry
is not in itself an issue but it does make things more complicated
than necessary.
Moreover, \ocpi is a push/pull protocol, so there are situations where
the \cpo connects as client to the \emsp, and situations where the
\emsp connects as client to the \cpo. Therefore, due to this asymmetry,
both sides of an \ocpi implementation need a static credential
that the other side can verify, both sides need a valid TLS server
certificate, and both sides need to implement authentication both on
the transport layer and on the application layer.


\medskip
\subsubsection*{Security improvement: replacing static credentials
with TLS client certificates for client authentication}
\label{sec:distls}
The use of static credentials is vulnerable to them leaking, and we do
not consider this to satisfy SR~2b.
In contrast, an authentication mechanism based on TLS certificates
doesn't require any secret static credential to be transmitted.
TLS is often already used to secure data in transit, which we will
discuss in Section~\ref{sec:tlsspec}, and to authenticate
the servers using server certificates for all protocols.
As all protocols are over TCP/IP, this is the natural choice.

\openadr, \oicp, and
\ocpp~2.0 in its highest security profile
already specify mandatory use of mutual TLS
authentication using client \& server certificates.
In these settings, the client certificate carries all the
information needed to identify and authenticate the communicating
party. After the TLS stack authenticates the party, it can then simply
pass the identification data up to the application layer, which can
then trust that that is the party that is being communicated with.
This takes the burden of authentication away from the application.

However, in the other protocols, this is currently not possible.
\ocpi even explicitly states client certificates are not used,
indicating that it was considered and decided against.
However, we do not believe that the credentials-based approach described
earlier in this Section provides anything that the client certificate
approach does not. The credentials are not used as cryptographic keys
or to provide any other means of security on the application layer;
they merely serve as secret identifiers. Such identifiers
could just as well be embedded in the TLS certificates.
Since every \cpo and every \emsp already needs to be part of a PKI,
needs to deal with server certificates, and needs some mechanism to
update its application-layer credentials,
we think that requiring the presence of client certificates is just as
reasonable as requiring the presence of static application-layer credentials.
We could therefore replace the static credentials in almost all protocols
with TLS client certificates, and make this mutual authentication
mandatory.

ISO~15118 is the only exception where we cannot use TLS client
certificates. Although EV contract certificates, which ISO~15118
uses to encode contract relations between EVs and \emsp{}s,
could at first glance be used as TLS client certificates by the EV,
ISO~15118 does not guarantee the presence of contract certificates.
Therefore, they cannot be relied upon to \emph{always} be used for TLS
client authentication, and some additional mechanism not using them
would be needed.
As such, it would make little sense to use TLS client authentication
even when such a certificate is present, since an attacker
could always claim not to have a certificate yet. Authentication
must therefore be performed on the application layer, e.g. using a
signature mechanism on the messages that require it.

Using client certificates has the additional benefit of significantly
shrinking the attack surface of an implementation.
Consider the case where an attacker of type 3, i.e. a network
attacker, without authentic credentials, tries to connect to a
system as a client. If authentication on the
application layer is used, then the application itself has to validate
the credential carried in the message. This exposes more code to malicious
input than if connection to the service depends on the presentation of
a valid client certificate, and the certificate check is done on the
transport layer before
permitting any application data handling. Authentication done on a
lower layer of the protocol stack effectively means that the higher
layers are no longer in the trusted computing base.
Effectively, any potentially
exploitable bugs in the application's message handling code are
shielded by the TLS authentication.
Of course any exploitable bugs in
the TLS authentication code are now a problem, but that
trusted computing base is
probably better
scrutinized than the rest of the application.

\subsection{Security of the transport links}
\label{sec:tlsspec}
The protocols that rely on TLS for server authentication also rely on
it to provide authenticity and confidentiality of the transport layer.
There are two major issues we see in this context:
\begin{enumerate}
	\item If TLS is not mandatory, implementers may choose not to use
		it at all.
	\item Even if TLS is mandatory, there are a lot of choices:
		\begin{itemize}
			\item which TLS versions are supported (\ocpi even refers
				to TLS as SSL),
			\item which cipher suites are mandatory, optional, or even
				prohibited,
			\item which certificate options are used,
			\item interpretations of what constitutes a valid certificate,
			\item \ldots
		\end{itemize}
\end{enumerate}
SR~3 seeks to solve the first issue by simply making TLS mandatory for
all protocols.

However, simply saying ``use TLS'' is not sufficient, because that
leaves the choices to the individual implementers, which does not help
with interoperability or security of the deployed systems.
\ocpi, \ochp, and \ocpp~1.5 and 1.6 provide no guidance for these
choices. \oicp in its available documentation only has a brief mention
of that client certificates are used to authenticate the clients,
without going into details on the TLS usage of the protocol.
In contrast, ISO~15118, \ocpp~2.0, and \openadr
have extensive descriptions of the TLS options and rationale for
the choices made. In effect, the protocol designers have already made
the choices that will ensure security on the transport layer, leaving
as little choice as possible to the implementer.

We believe that a strict specification based on security analyses is
preferable to a loose or barely existent specification left to the
implementer, and only a standard that makes TLS mandatory \emph{and}
specifies how to use TLS satisfies SR~3.

\medskip
\subsubsection*{Security improvement: complete specification and
unification of TLS and the PKI}
\label{sec:distlsspec}
To simplify the ecosystem and, by extension, lower the chance of
interoperability bugs and security issues resulting from those,
ideally all protocols would use unified TLS requirements. Furthermore,
since all protocols need some form of PKI for their TLS functionality,
it would be desirable to have a single unified PKI that can fulfil all
the certificate requirements of the EV-charging ecosystem. ISO~15118
and \openadr have some requirements and limitations on their
certificates and, by extension, their PKI. A report by ElaadNL explains
the TLS PKI as required by ISO~15118 for implementers \cite{elaadpki}.  We will look
at the technical and and organizational details of unifying the TLS
requirements and PKI for all these protocols in a separate
publication, but we can summarize our main findings here.

From a technical point of view, unifying TLS requirements is simple.
ISO~15118 and \ocpp~2.0 already have strict rules on their allowed TLS
cipher suites, and the only common cipher suite is
TLS\-\_\-ECDHE\-\_\-ECDSA\-\_\-WITH\-\_\-AES\_128\_GCM\-\_\-SHA256 from
TLS 1.2.
This is a state-of-the-art cipher suite, however, which is also still
available in TLS~1.3 and therefore future-proof.
We do not see a good reason to opt for more configurability.
However, it might be desirable to add another cipher suite that is
available in both TLS~1.2 and TLS~1.3 but which is built on different
primitives. This would ensure that the ecosystem can remain secure
should the current cipher suite be broken, until all systems in the
field can be updated to newer cipher suites.
A separate standard specifying these requirements, that other
protocols can then refer to, could be used. This also makes it easier
to update the requirements if vulnerabilities are found. An example of
how this could look is the chapter on TLS in~\cite{enisa-protocols}.

Similarly, from a technical point of view, using a single unified PKI
should be possible. Although ISO~15118 has very extensive technical
requirements on its certificates, these do not necessarily clash with
the requirements that other protocols have. Even if technical
requirements turn out to be incompatible, a unified PKI could simply
have different trees for different protocols under the same root CA.

However, even though it seems that there are no major technical issues
blocking such a unification, in the years since publication of
ISO~15118 there has been little move towards establishing a PKI for
it, let alone unifying requirements with the other protocols. The
exact reasons for this are unclear, but
we have noticed some reluctance from the market to be tied
down to a unified PKI. There are multiple ways to organize such a PKI,
and market parties are currently exploring possible setups. The
clearing house Hubject, the organization behind \oicp, is already
running a PKI for use in \oicp; but since they are a clearing house
they have an interest in the EV market itself. ElaadNL is piloting a
few different technical options to tie multiple PKIs together.
We believe that the best option is an independent certificate
authority, not tied to a market player, which is overseen by an
independent or at least cross-organizational body to ensure a fair and
open market.

\subsection{Lack of (end-to-end) security on the application layer}
\label{sec:e2e}
As explained in Section~\ref{sec:securityrequirements} where we
discussed SR~4a and SR~4b,
TLS cannot provide end-to-end security where parties forward data in
transit, nor security for data at rest. Another
mechanism on the application layer
is required to satisfy SR~4a, SR~4b, and SR~5.
Of all protocols we have listed, only ISO~15118 and \ocpp~2.0 currently
provide such a mechanism. ISO~15118 provides XML signatures and
partial encryption of a select subset of its messages. \ocpp~2.0
provides optional message signing for entire \ocpp messages.
It seems that the other protocols have not considered
end-to-end security as a goal.

\medskip
\subsubsection*{Security improvement: security on the application layer}
\label{sec:dise2e}
%
As an initial improvement, at the very least, all protocols should
ensure that digital signatures added as part of ISO~15118 and \ocpp~2.0
are forwarded along with the data, and are still verifiable:
there is the risk that changing the data format, notably
from XML to JSON, will mean the signatures over the original data
can no longer be verified if data that was originally part of a
signed packet is discarded.

However, we believe it is possible to go further.
All protocols should be able to satisfy SR~4a, 4b, and 5.
The mechanisms for satisfying these SRs in ISO~15118
are only applied to a select subset of its messages, and are not
applicable to the other protocols due to their implementation using
XML signatures.
We would like to see a more generic solution that is
relatively easy to apply to all communication. The mechanism in
\ocpp~2.0 to satisfy SR~4a and SR~5 is applicable to all its messages, 
but signs the entire payload of a message at once. Although this could
be fairly easily applied to any other JSON-based protocol, it does not
satisfy SR~4b. Furthermore, the practice of signing
entire messages at once conflicts with requirements from the GDPR,
as will be explained in Section~\ref{sec:secreqsgdpr}.
In light of this, we have proposed a different security scheme in
\cite{vanaubelrijneveldpoll2019} that would provide both end-to-end
security for data in transit, and authenticity and non-repudiation for
data at rest.

ISO~15118 requires compatible charge points and
cars, as well as a running contract. Since cars not implementing
ISO~15118 will be around for decades, External Identification
Means (EIM) with e.g. RFID cards or EMV cards, as discussed in
Section~\ref{sec:disevdriverauth}, will remain for the foreseeable future.
In that case the car cannot sign data.
One way to achieve a comparable level of trust is to have the EIM used
sign the data instead.
E.g. if a custom RFID solution or a smartphone app is used for driver
authentication, as mentioned in Section~\ref{sec:disevdriverauth},
these could be provisioned with some key material that is used to
sign the final meter reading when the driver ends the charge
session and unlocks the charge cable from the car, effectively
implementing the most important security features from ISO~15118 on
the card.
If this is not possible, only the charge point could sign data.
However, this is strictly weaker than the car or EIM signing it, since
the charge point is under management of the \cpo, not the EV driver.
Therefore, the \cpo would not have as strong a case if the EV driver
decided to dispute a transaction.

\section{Privacy of the EV driver}
\label{sec:privacy}
In our security analysis we have largely ignored the privacy issues of
the EV-charging ecosystem, but we do believe that there are pressing
privacy issues that the industry needs to deal with, e.g. as described
in \cite{langer2013}.
A lot of the data being exchanged is personal data
under the GDPR \cite{GDPR}. This does not mean the processing cannot
happen, but it does mean certain requirements need to be met.

One of the most important requirements of the GDPR is that only data
required for a specific purpose is processed, and only by those
parties that actually need to process it. Processing is broadly
defined and includes transmission, storage, and deletion. This clashes
with the current setup of the EV-charging ecosystem because the
proxying \cpo{}s see data pass in plaintext. This is one reason for
SR~4b, the end-to-end confidentiality requirements.

\subsection{Privacy Impact Assessments for the industry}
The following are some privacy-related highlights that drew our
attention during the security analysis for this paper:
\begin{itemize}
	\item ISO~15118 states as a requirement that private information
		shall only be readable by the intended recipient, and be
		transferred only when necessary. It goes on to equate
		confidentiality with privacy, which is a very narrow view on
		privacy. It has no additional comments on what constitutes
		``private information'', it ignores the issue of deciding in
		the first place what information is required by each actor,
		and it does not consider the additional information that could
		be derived from that data by the recipient at all.

	\item \ocpi does mention that contract IDs are linked to persons
		and therefore the user should be aware of privacy issues. But
		it also phrases the handling of CDRs as follows \cite{OCPI}:
		\begin{quotation}
			``A \cpo is not required to send \textit{all} CDRs to
			\textit{all} \emsp{}s, it is allowed to only send CDRs to the
			\emsp that a CDR is relevant to.''
		\end{quotation}
		The first part of this phrase implies it would be acceptable
		to send CDRs to other parties than the \emsp that an EV driver
		has a contract with.  Since a CDR contains everything required
		for billing, it necessarily contains personal data: location,
		time of charge, amount of energy charged.
		As such, sending a CDR to any \emsp \emph{other} than the one
		it is relevant to is a violation of the GDPR.

	\item \ocpp~2.0 can retrieve and remove customer
		information from a charge point ``for example to be compliant
		with local privacy laws'' \cite{OCPP}. Although this seems
		to be to ensure that charge points can facilitate GDPR
		requirements, it is the only time privacy is mentioned in
		\ocpp.
\end{itemize}
It seems that the individual parties are aware of the potential for
privacy issues, but nobody so far has really looked at all the data
that all these protocols are supposed to exchange and figure out what
data is really required, by whom, for what purposes, and for how long.
We have spoken to individual actors who have performed Privacy
Impact Assessments (PIAs) on their own practices. But these PIAs do not
necessarily lead to a privacy-friendly ecosystem. For that,
the EV-charging ecosystem needs a standardization effort to determine
precisely what data needs to be exchanged between which roles.
This goes beyond a PIA of a single
actor: the concerns cut across all the \cpo{}s, \emsp{}s, car
manufacturers, value-added service providers, and all other actors
that make up this ecosystem.

We would propose to solve this in a way similar to how the Dutch smart
metering ecosystem has \cite{vanaubelpoll2019}:
the actors that fill a certain role
organize in an industry consortium, and determine what data they
actually need to provide their services. This effort would involve
consortium-wide PIAs, and should result in shared codes of
conduct that cover the use of personal data of actors in each role.
At a minimum this would result in codes of conduct for \cpo{}s and
\emsp{}s. Then, \ocpp, \ocpi, and other affected
protocols would be updated to implement the codes of conduct; in
particular, protocols should ensure that data that is not required
by an actor is not mandatory in messages to that actor.

\subsection{Security Requirements versus the GDPR}
\label{sec:secreqsgdpr}
As mentioned in Section~\ref{sec:securityrequirements}, the
requirements of the GDPR could clash with SR~4a, SR~4b, and SR~5.
One of these requirements is that data is removed as soon as
it is no longer needed. 
Suppose we have a CDR that contains, among other things, a customer
identifier, location, time, total cost of charge session, and amount
of energy charged. After billing the EV driver, the location may no
longer be relevant, in which case it should be removed. However, the
rest of the CDR, especially total cost of the session, may need to be
kept. A signature over an entire CDR usually requires that entire CDR
for verification, so then the location cannot be removed without
invalidating the signature that also proves authenticity of the cost
of the session.

In a similar way, such a plain signature mechanism would clash with
the aforementioned requirement that data is only processed by those
parties that need to process it.
The messages that are received and forwarded by \cpo{}s often carry
information only intended only for the \cpo, which should not be
forwarded to the \emsp. Consider the example of the CDR again:
arguably the \emsp does not need the location information of the
customer \emph{at all}. In the current setup it is possible to drop
the location from the message that is sent to the \emsp, but with a
plain signature mechanism over the entire CDR, nothing can be
selectively removed without invalidating the signature.

The end-to-end security scheme introduced in
\cite{vanaubelrijneveldpoll2019} can be used to satisfy SR~4a, SR~4b,
and SR~5, while also enabling the user to satisfy the requirements
from the GDPR.

\section{Future work}
\label{sec:futurework}
In the current ecosystem, charge points require a network connection
to communicate with the back-end systems of the \cpo. This network
connection may not be reliable, which is one of the reasons for SR~1c:
an EV driver should be able to charge even if the charge point is
offline.

The options discussed in \ref{sec:disevdriverauth} deal with the
case of driver authentication for post-fact billing. However, 
performing payments at the charge point itself, or by online
transaction, is also possible in \ocpp~2.0. In such a case, no
additional authentication is required;
all that is needed is that the charge point is able to verify that a
transaction was performed. However, this does require the charge point
to be online, potentially violating SR~1c.

Another reason that \ocpp~2.0 facilitates starting the charge session
directly from the \cpo back-end system is the potential to use a
smartphone app to start charging.
This also requires the charge point to have an online
network connection to receive the relevant start commands from the
\cpo, but more importantly, it requires the smartphone to have an
online network connection to send the start command from the app to
the \cpo.
As suggested in Section~\ref{sec:disevdriverauth}, NFC-capable
smartphones could be used for driver authentication to the charge point,
which would require a reader on the charge point capable of
communicating with the phone.
Combining these concepts, it would be possible to use the NFC-capable
smartphone's network connection to proxy the communication between
\cpo and charge point. Instead of sending the start command directly to
the charge point, the \cpo sends a signed session description to
the smartphone, which in turn sends it via NFC to the charge point.
If SR~4a, end-to-end authenticity, is satisfied, the charge point can
check the validity of this description, trust that the session is
legitimate, and charge the car accordingly.
The (security) details of such a mechanism could be explored in future
work.

\section{Conclusions}
\label{sec:conclusion}
Our primary conclusion is that, although the EV-charging ecosystem is
showing a promising move towards using TLS for authentication and for
secure communication links everywhere, this is insufficient, as
explained in Section~\ref{sec:securityrequirements}. The
ecosystem needs end-to-end security for data in transit, and long-term
authenticity and non-repudiation for data at rest, neither of which
can be provided by TLS. This is required so that actors do
not need to blindly trust one another. Data in transit needs to be
secured not just against attackers listening in on the network
traffic, but also against the proxying parties such as charge points,
charge point operators, and clearing houses. Data at rest needs to
provide some guarantees: an \emsp should be able to prove that a \cpo
really did send a certain Charge Detail Record, and all parties should
be able to verify that such a Charge Detail Record was not tampered
with.
We believe that it is feasible to add end-to-end, long-term authenticity
\emph{and} end-to-end confidentiality to all data exchanged,
while taking into account privacy issues and GDPR
compliance, as explained in Sections~\ref{sec:dise2e}
and~\ref{sec:privacy}.
The ability of ISO~15118 for the car to sign meter readings is a first
step towards this, but is highly specific and not applicable
to the other protocols. A potential solution is
given in \cite{vanaubelrijneveldpoll2019}.

We see some other pressing security issues in the current versions of the
protocols in use:
\begin{enumerate}
	\item TLS is not yet mandatory. This is the bare minimum of
		security, as it is needed to protect the individual
		communication links against attackers reading and modifying
		the network traffic. Where TLS is mandatory, it is often
		underspecified. Ideally, the ecosystem would work towards a
		single TLS specification and public key infrastructure, which
		could then be adopted by all protocols, as described in
		Section~\ref{sec:distlsspec}. We intend to explore this in
		future work.

	\item Several protocols use a weak form of authentication
		between systems, which we explained in
		Section~\ref{sec:systemauth}.
		Using TLS with client certificates solves that issue.
		\ocpp~2.0, \oicp, and \openadr demonstrate the best current
		practice w.r.t. using client certificates, with \ocpp~2.0 being
		the most extensive in its specification of how TLS and client
		certificates should be used.  This only solves authentication
		between directly communicating parties, however -- proxied
		communication is not authenticated.
	
	\item Authentication of the EV driver is weak, based solely on RFID
		UIDs.  ISO~15118's specification of contract certificates and
		the authentication method Plug-and-Charge is stronger.
		Unfortunately, legacy EVs that do not implement ISO~15118 or
		Plug-and-Charge will remain for a long time, so even though a better
		authentication system could be established, support for the legacy
		RFID systems will need to remain for the foreseeable future.
		However, that does not preclude these RFID systems from being
		improved, as discussed in~\ref{sec:disevdriverauth}.

		The EV-charging ecosystem is not the first to have this
		problem. The banking sector and the public transport sector
		have both built solutions to deal with cross-party
		authentication. It would be beneficial to explore how
		applicable their solutions are to this ecosystem.
\end{enumerate}

Finally, though not directly related to the security concerns at the
focus of this paper, we wish to draw attention to the fact that not
all the protocols are well-aligned with the current market.
This is particularly the case for \oscp and \openadr. These protocols
aim to offer a DSO more flexibility in congestion management.  This is
clearly in the interest of the DSO: making better use of fixed
capacity might reduce the required investments in distribution
infrastructure. However, if \cpo{}s always have contracts
for a fixed capacity, there is no way for DSOs to pass on this
economic advantage to \cpo{}s, and hence no economic incentive for \cpo{}s
to use such flexibility -- for them, the cost of the network is an
externality.
This leads to a typical `tragedy of the commons', where the market
forces lead to a sub-optimal solution for society as a whole.
This is an interesting parallel, in that
economic disincentives are also notorious as a root cause of cyber
security issues \cite{Herley2009}. This may well turn out to be
there case here: for some parties in the EV market it may be against
their short term individual economic interests to invest in cyber
security, an investment that would come at the expense of e.g. price
or quickly building up market share.



\section*{Acknowledgements}
Funding: This work was supported by the EU Regional Development
Fund (ERDF), as part of the project Charge \& Go.

\bibliographystyle{elsarticle-num}
\bibliography{EVcharging}

\end{document}